\documentclass[groupedaddress,
  aip,
  pop,
  amsmath,amssymb,
 reprint,
]{revtex4-1}
\usepackage{graphicx,color}
\usepackage{amsmath}
\usepackage{natbib}
\usepackage{epsfig}
\begin{document}

\title{Unified description of sound velocities in strongly coupled Yukawa systems of different spatial dimensionality}

\author{Sergey A. Khrapak}
\email{Sergey.Khrapak@dlr.de}
\affiliation{Institut f\"ur Materialphysik im Weltraum, Deutsches Zentrum f\"ur Luft- und Raumfahrt (DLR), 82234 We{\ss}ling, Germany;\\ Joint Institute for High Temperatures, Russian Academy of Sciences, 125412 Moscow, Russia}

\date{\today}

\begin{abstract}
Sound velocities in classical single-component fluids with Yukawa (screened Coulomb) interactions are systematically evaluated and analyzed in one-, two-, and three spatial dimensions (${\mathcal D}=1,2,3$). In the strongly coupled regime the convenient  sound velocity scale is given by $\sqrt{Q^2/\Delta m}$, where $Q$ is the particle charge, $m$ is the particle mass, $n$ is the particle density, and $\Delta=n^{-1/{\mathcal D}}$ is the unified interparticle distance. The sound velocity can be expressed as a product of this scaling factor and a dimension-dependent function of the screening parameter, $\kappa=\Delta/\lambda$, where $\lambda$ is the screening length. A unified approach is used to derive explicit expressions for these dimension-dependent functions in the weakly screened regime ($\kappa\lesssim 3$). It is also demonstrated that for stronger screening ($\kappa\gtrsim 3$), the effect of spatial dimensionality virtually disappears, the longitudinal sound velocities approach a common asymptote, and a one-dimensional nearest-neighbor approximation provides a relatively good estimate for this asymptote. This result is not specific to the Yukawa potential, but equally applies to other classical systems with steep repulsive interactions. An emerging relation to a popular simple freezing indicator is briefly discussed. Overall, the results can be useful when Yukawa interactions are relevant, in particular, in the context of complex (dusty) plasmas and colloidal suspensions.       
     
\end{abstract}

\maketitle

\section{Introduction}

Investigations into linear and non-linear waves in complex (dusty) plasmas -- systems of charged macroscopic particles immersed in a plasma environment -- is an active research area with many interesting topics, such as e.g. sound (dust-acoustic) waves, instabilities, Mach cones, shocks, solitons, and turbulence.~\cite{MerlinoPoP1998,FortovPR,ShuklaRMP2009,Merlino2014,ThomasPPCF2018}  In experiments, sufficiently long wavelengths are usually easy accessible for investigation, which exceed considerably the characteristic interparticle separation. At these wavelengths collective excitations exhibit acoustic-like dispersion and the sound velocities play central role in characterizing the system.

The particle charge in complex plasmas is typically very high ($10^3-10^4$ elementary charges for a micron-range sized particles). 
Due to strong electrical repulsion between the particles they usually form condensed liquid and solid phases. It is well understood that dispersion properties of strongly coupled complex plasmas significantly deviate from those characteristic for an ideal gaseous plasma.~\cite{FortovUFN,FortovPR,Bonitz2010,DonkoJPCM2008} Strong coupling effects affect the magnitudes of sound velocities.~\cite{KalmanPRL2000,KhrapakPPCF2015} Strongly coupled complex plasma fluids in two and three dimensions can support transverse excitations at finite (sufficiently short) wavelengths.~\cite{OhtaPRL2000,PramanikPRL2002,NosenkoPRL2006} Instability thresholds (e.g. of the ion current instability) are shifted at strong coupling.~\cite{RosenbergPRE2014}  

Waves in complex plasmas are investigated in one-dimensinal (1D), two-dimensional (2D), and three-dimensional configurations (3D).
1D linear particle arrangements as well as 1D and quasi-1D particle rings are formed by creating appropriate confining potential configurations above the negatively charged surface (electrode), responsible for particle levitation.~\cite{PetersPLA1996,HomannPRE1997,SheridanPoP2016} 2D and quasi-2D layers are extensively studied in laboratory experiments with radio-frequency (rf) discharges, where the levitating particles form horizontal layer(s) in the plasma sheath above the lower rf electrode.~\cite{Zuzic1996,PieperPRL1996,NunomuraPRL2002,NunomuraPRE2002,NunomuraPRL2005} Waves in large 3D particle clouds have been initially observed in a Q-machine,~\cite{BarkanPoP1995} and then in dusty plasmas formed in a positive column (sometimes stratified) of direct-current glow discharges,~\cite{MolotkovJETP1999,FortovPoP2000,RatynskaiaPRL2004,KhrapakPRE2005,EThomasCPP2009} as well as in various experiments under microgravity conditions.~\cite{KhrapakPoP2003,YaroshenkoPRE2004,PielPRL2006,MenzelPRL2010,HimpelPoP2014,YaroshenkoPoP2019}  

Sound velocities can be relatively easy and accurately measured in experiments~\cite{NunomuraPRE2002,NunomuraPRL2005,KhrapakPoP2003,SaitouPRL2012} and contain important information about the systems investigated. 
 
The purpose of this paper is to provide a unified description of sound velocities in strongly coupled complex plasmas in 1D, 2D, and 3D geometries. It is assumed that the particles are interacting via the isotropic pairwise Yukawa (screened Coulomb) potential. Simple practical formulas are obtained, which are applicable to condensed fluid and solid phases. In particular, it is demonstrated that the sound velocities are given by the product of the relevant velocity scale  $\sqrt{Q^2/\Delta m}$ and the screening function $f(\kappa)$, where $Q$ is the particle charge, $\Delta=n^{-1/{\mathcal D}}$ is the characteristic interparticle separation, $n$ is the density, ${\mathcal D}$ is the dimensionality, $m$ is the particle mass, and $\kappa$ is the screening parameter defined as the ratio of the interparticle separation to the screening length $\lambda$,  that is $\kappa=\Delta/\lambda$.  The properties of $f(\kappa)$ in 1D, 2D, and 3D cases are investigated. In particular, the two regimes of weakly screened ($\kappa\ll 1$) and strongly screened interactions ($\kappa\gg 1$) are considered in detail. Important consequences and relations are discussed.      

Yukawa systems are characterized by the repulsive interaction potential of the form $\phi(r)=(Q^2/r)\exp(-r/\lambda)$.  Regardless of dimensionality, the phase state of the system is conventionally described  by the two dimensionless parameters, which are the (Coulomb) coupling parameter $\Gamma=Q^2/\Delta T$, and the screening parameter $\kappa$, where $T$ is the system temperature (in energy units, so that $k_{\rm B}=1$). It is important to note that very often the Wigner-Seitz radius is used as a length unit, instead of $\Delta$. The Wigner-Seitz radius is defined from $4\pi n a^3/3=1$ in 3D, $\pi a^2 n =1$ in 2D, and $na=1$ in 1D (that is only in 1D we have $\Delta=a$). Correspondingly, $\Gamma$ and $\kappa$ are also often defined in terms of $a$ and one should pay attention to this. In this paper $\Delta$ is exclusively used as a length unit. 

The Yukawa potential is considered as a reasonable starting point to model interactions in complex (dusty) plasmas and colloidal dispersions,~\cite{FortovPR,IvlevBook} although in many cases the actual interactions (in particular, their long-range asymptotes) are much more complex.~\cite{TsytovichUFN1997,FortovPR,KhrapakCPP2009,RatynskaiaPoP2006,KhrapakPRL2008,KhrapakPoP2010,ChaudhuriIEEE2010,ChaudhuriSM2011,LampePoP2015} This is particularly true in cases when electric fields and ion drifts are present, resulting in plasma wakes and wake-mediated interactions.~\cite{VladimirovPRE1995,KompaneetsPoP2009,HutchinsonPoP2011,LudwigNJP2012,KompaneetsPRE2016} 
The sound velocities will be certainly affected by deviations from the assumed Yukawa potential, but we do not attempt to discuss this issue here. Recently, the effect of long-range deviations from the pure Yukawa potential on the dispersion relations of the longitudinal waves in isotropic complex plasmas have been investigated.~\cite{Fingerprints}  The behavior of waves in a 1D dusty plasma lattice where the dust particles interact via Yukawa plus electric dipole interactions has been theoretically studied in Refs.~\onlinecite{YaroshenkoNJP2006,RosenbergJPP2015}.

The paper is organized as follows. In section \ref{sec_sound} the unified approach to the calculation of sound velocities in strongly coupled Yukawa systems in 1D, 2D, and 3D is presented. Main results are summarized in Section~\ref{MainResults}. Here the weakly screened regime is analyzed in detail. Approximate expressions for the sound velocities in systems with steeply repulsive potentials are derived, and it is explained why spatial dimensionality does not affect considerably the magnitude of sound velocities in this regime.  This is followed by conclusion in Sec.~\ref{Concl}. Relation to a simple freezing indicator of classical 3D fluids proposed earlier is then briefly discussed in Appendix~\ref{freezing}.  

\section{Sound velocities in different spatial dimensions}\label{sec_sound}

Strongly coupled Yukawa systems support one longitudinal mode in 1D case, one longitudinal and one transverse mode in 2D case, and one longitudinal and two transverse modes in 3D case.

The longitudinal sound velocities can be obtained from the conventional hydrodynamic (fluid) approach.~\cite{LL_hydrodynamics} This requires knowledge of an appropriate equation of state. The standard adiabatic sound velocity is $c_{\rm s}=\sqrt{(1/m)(\partial P/{\partial n})_s}$, where $P$ is the pressure of a {\it single component} Yukawa system and the subscript $s$ denotes that the derivative with respect to density is taken at constant entropy.  Note that  $(\partial P/{\partial n})_s=\gamma(\partial P/{\partial n})_T$, where $\gamma = c_{\rm p}/c_{\rm v}$ is the adiabatic index. For strongly coupled Yukawa systems we have $\gamma\simeq 1$, which is a general property of soft repulsive interactions.~\cite{KhrapakPRE2015_Sound,SemenovPoP2015,FengPoP2018} This fluid approach has been exploited previously for Yukawa systems in 3D case~\cite{KhrapakPPCF2015,KhrapakPRE2015_Sound} as well as in 2D case.~\cite{SemenovPoP2015} Generalization to 1D case is trivial.    

The sound velocities of strongly coupled Yukawa systems can also be obtained from infinite-frequency (instantaneous) elastic moduli, directly related to the instantaneous normal modes.~\cite{Stratt1997,KhrapakSciRep2017,WangPRE2019} This approach is applicable to fluids and solids and allows to calculate both the longitudinal and transverse sound velocities in a universal manner and hence is adopted here.    

The elastic waves modes (instantaneous normal modes) in the strongly coupled plasma-related fluids are rather well described by the quasilocalized charge approximation (QLCA),~\cite{GoldenPoP2000,KalmanPRL2000,DonkoJPCM2008} also known as the quasi-crystalline approximation (QCA).~\cite{Hubbard1969,Takeno1971,KhrapakSciRep2017} This approximation relates wave dispersion relations to the interparticle interaction potential $\phi(r)$ and the equilibrium radial distribution function (RDF) $g(r)$, characterizing structural properties of the system. It can be considered as either a generalization of the random phase approximation or as a generalization of the phonon theory of solids.~\cite{Hubbard1969} The latter point of view is particularly relevant, because in the special case of a cold crystalline solid the QCA dispersion reduces to the ordinary phonon dispersion relation,~\cite{Hubbard1969} justifying the approach name. It is known that for 2D Yukawa systems, the angularly averaged lattice dispersions are remarkably similar to the isotropic QCA fluid dispersions.~\cite{SullivanJPA2006,HartmannIEEE2007} It is not very unreasonable to expect similar behavior in the 3D case. 

The long-wavelength limits of the QCA dispersion relations can be used to define the elastic longitudinal and transverse sound velocities, $c_l$ and $c_t$, as explained in detail below.  The relation to the thermodynamic (adiabatic $\simeq$ isthermal) sound velocity is then
$c_s^2\simeq c_l^2-(4/3)c_t^2$ in 3D and $c_s^2=c_l^2-c_t^2$ in 2D. For Yukawa interactions (as well as for other soft long-ranged repulsive interaction potentials) the strong inequality $c_l^2\gg c_t^2$ holds at strong coupling. This implies that we have approximately $c_s\simeq c_l$. The accuracy of this relation has been numerously tested for strongly coupled Yukawa fluids,~\cite{KhrapakPRE2015_Sound,KhrapakPPCF2015,KhrapakPoP2016_Relations} as well as other soft interactions,~\cite{KhrapakSciRep2017,GoldenPRE2010,KhrapakPoP2016_Log} both in 3D and 2D cases.       
  
The general QCA (QLCA) expressions for the longitudinal and transverse dispersion relations are 
\begin{equation}\label{omegaL}
\omega_{l}^2=\frac{n}{m}\int\frac{\partial^2 \phi(r)}{\partial z^2}g(r)\left[1-\cos(\bf{kz})\right]d{\bf r},
\end{equation} 
and 
\begin{equation}\label{omegaT}
\omega_{t}^2=\frac{n}{m}\int\frac{\partial^2 \phi(r)}{\partial x^2}g(r)\left[1-\cos(\bf{kz})\right]d{\bf r},
\end{equation} 
where $\omega$ is the frequency and ${\bf k}$ is the wave vector. It is worth mentioning at this point that $\omega_l^2$ and $\omega_t^2$ can be identified as the potential (excess) contributions to the normalized second frequency moments of the longitudinal and transverse current spectra, $C_{l/t}(k,\omega)$.~\cite{BulacaniBook} Kinetic terms, which are absent in the QCA approach [$3(T/m)k^2$ for the longitudinal branch and  $(T/m)k^2$ for the transverse one], are relatively small at strong coupling. Thus, the formal essence of the QCA approach is just to approximate the actual dispersion relations by the excess contributions to the second frequency moments of the corresponding current spectra.   

We proceed further as follows.  
The derivatives of the pair interaction potential in Eqs.~(\ref{omegaL}) and (\ref{omegaT}) are evaluated from
\begin{displaymath}
\frac{\partial^2 \phi(r)}{\partial x_{\alpha}^2} = \phi''(r)\frac{x_{\alpha}^2}{r^2}+\frac{\phi(r)'}{r}\left(1-\frac{x_{\alpha}^2}{r^2}\right),
\end{displaymath} 
where $x_{\alpha}=x,y,z$ in 3D,   $x_{\alpha}=x,z$ in 2D, $x_{\alpha}=z$ in 1D, and $r=\sqrt{\sum_{\alpha}x_{\alpha}^2}$. Note also that from symmetry 
\begin{displaymath}
\frac{\partial^2 \phi(r)}{\partial x^2}=\frac{\partial^2 \phi(r)}{\partial y^2}=\frac{1}{2}\left[\Delta \phi(r) -\frac{\partial^2 \phi(r)}{\partial z^2}\right]
\end{displaymath}
in 3D, and
\begin{displaymath}
\frac{\partial^2 \phi(r)}{\partial x^2}=\Delta \phi(r) -\frac{\partial^2 \phi(r)}{\partial z^2}
\end{displaymath}  
in 2D.

Let us consider isotropic fluids with pairwise interactions of the form
\begin{equation}
\phi(r)=\epsilon f (r/\sigma),
\end{equation} 
where $\epsilon$ is the energy scale and $\sigma$ is the length scale. Except for some special cases (in the present context this corresponds to the unscreened Coulomb interaction limit, which will not be considered explicitly), the long-wavelength dispersion is acoustic:
\begin{equation}
\lim_{k\rightarrow 0}\frac{\omega_{l}^2}{k^2}=c_{l}^2, \quad\quad \lim_{k\rightarrow 0}\frac{\omega_{t}^2}{k^2}=c_{t}^2.
\end{equation}
The emerging elastic longitudinal and transverse sound velocities can be presented in a universal form~\cite{KhrapakPoP2016_Relations}    
\begin{equation}\label{disp_gen_3D}
c_{{l}/{t}}^2=\omega_{{\mathcal D}}^2\sigma^2\int_0^{\infty}dx x^{{\mathcal D}+1} g(x) \left[{\mathcal A}\frac{f'(x)}{x}+{\mathcal B}f''(x)\right],  
\end{equation}
where $x=r/\sigma$ is the reduced distance. The ${\mathcal D}$-dimensional effective frequencies $\omega_{\mathcal D}$ and the coefficients ${\mathcal A}$ and ${\mathcal B}$ are summarized in Table~\ref{Tab1}. The last line in Table ~\ref{Tab1} simply reflects the fact that the transverse mode is absent in 1D case and the integration over the positive and negative parts of $z$-axis is equivalent to the doubled integration over the positive part.

An important remark about the transverse dispersion relation in fluids should be made at this point.  Although strongly coupled (dense) fluids do support the transverse waves propagation, their dispersion is somewhat different from that in a solid. The existence of transverse modes in fluids is a consequence of the fact that their response to high-frequency short-wavelength perturbations is similar to that of a solid.~\cite{ZwanzigJCP1965} However, shear waves in fluids cannot exist for arbitrary long wavelengths. The minimum threshold wave number $k_*$ emerges, below which transverse waves cannot propagate. This phenomenon, often referred to as the $k$-gap in the transverse mode, is a very well known property of the fluid state.~\cite{HansenBook,Trachenko2015}  Locating $k_*$ for various simple fluids in different parameter regimes and investigating $k$-gap consequences on the liquid state properties is an active area of research.~\cite{GoreePRE2012,YangPRL2017,KhrapakJCP2018,KhrapakJCP2019,KryuchkovSciRep2019} For our present purpose it is important that the inclination of the dispersion curve $\partial \omega_t/\partial k$ near the onset of the transverse mode at $k>k_*$ can be well approximated by $c_t$. Thus, the latter is a meaningful quantity both in solid and strongly coupled fluid states.        

\begin{table}
\caption{\label{Tab1} The coefficients ${\mathcal A}_{l/t}$ and ${\mathcal B}_{l/t}$ appearing in Eq.~(\ref{disp_gen_3D}) for the longitudinal ($l$) and transverse ($t$) sound velocities, as well as ${\mathcal D}$-dimensional nominal frequencies and the coefficients  ${\mathcal C}_{\mathcal D}$ in 3D, 2D, and 1D spatial dimensions.}
\begin{ruledtabular}
\begin{tabular}{lcccccc}
$\mathcal D$ & $\omega_{\mathcal D}^2$ &  ${\mathcal C}_{\mathcal D}$ &${\mathcal A}_{l}$ & ${\mathcal B}_{l}$ &  ${\mathcal A}_{t}$ & ${\mathcal B}_{t}$  \\ \hline
3D & $4\pi n\epsilon\sigma/m$ & $4\pi$ & $\frac{1}{15}$ & $\frac{1}{10}$ & $\frac{2}{15}$ & $\frac{1}{30}$    \\
2D & $2\pi\epsilon n/m$ & $2\pi$ & $\frac{1}{16}$ & $\frac{3}{16}$ & $\frac{3}{16}$ & $\frac{1}{16}$   \\
1D& $2\epsilon n/m\sigma $ & $2$ & 0 & $\frac{1}{2}$ & 0 & 0 \\
\end{tabular}
\end{ruledtabular}
\end{table}  

Next we take $\sigma=\Delta$ and assume Yukawa interaction potential between the particles. This implies $\epsilon=Q^2/\Delta$ and $f(x)=\exp(-\kappa x)/x$. The expressions for the longitudinal and transverse sound velocities become
\begin{equation}
\begin{split}
c_{{l}/{t}}^2={\mathcal C}_{{\mathcal D}}\left(\frac{Q^2}{\Delta m}\right)\int_0^{\infty}dx x^{{\mathcal D}-2}\exp(-\kappa x) g(x) \\ \left[{\mathcal B}_{l/t}\kappa^2 x^2 +(2{\mathcal B}_{l/t}-{\mathcal A}_{l/t})(1+\kappa x)\right].  
\end{split}
\end{equation}
The numerical coefficients ${\mathcal C}_{\mathcal D}$ are provided in Table~\ref{Tab1}. At this point it is also useful to introduce the universal velocity scale $c_0=\sqrt{Q^2/\Delta m}$. Note that $c_0=\sqrt{\Gamma}v_T$, where $v_T=\sqrt{T/m}$ is the thermal velocity. 

The excess internal (potential) energy can also be expressed using the RDF and the pair interaction potential. The expression for the excess energy per particle in units of temperature is~\cite{HansenBook}
\begin{equation}\label{energy}
u_{\rm ex}= \frac{n}{2T}\int d{\bf r} \phi(r)g(r).
\end{equation}   
For the Yukawa interaction potential in ${\mathcal D}$ dimensions this yields
\begin{equation}
u_{\rm ex}={\mathcal C}_{\mathcal D}\frac{\Gamma}{2}\int_0^{\infty}dx x^{{\mathcal D}-2}\exp(-\kappa x)g(x),
\end{equation}
where we have used the identity $\epsilon/T=Q^2/\Delta T\equiv \Gamma$. 

Finally, the following line of arguments is used. In the special case of a cold crytalline solid, the RDF represents a series of delta-correlated peaks corresponding to a given lattice structure. Assuming that the lattice structure is fixed (in fact, the equilibrium lattice structure changes from bcc to fcc when $\kappa$ increases~\cite{RobbinsJCP1988,HamaguchiPRE1997,VaulinaPRE2002} in 3D case, but this is not important for our present purpose) the RDF is a universal function of $x$: $g(x;\Gamma,\kappa)=g(x)$ (for simplicity we keep isotropic notation). Independence of $g(x)$ of $\kappa$ allows us make use of the following identities:
\begin{displaymath}
{\mathcal C}_{\mathcal D}\Gamma\int_0^{\infty}dx x^{{\mathcal D}-2}\exp(-\kappa x) g(x) = 2u_{\rm ex},
\end{displaymath}
\begin{displaymath}
{\mathcal C}_{\mathcal D}\Gamma\int_0^{\infty}dx x^{{\mathcal D}-2}\kappa x\exp(-\kappa x) g(x) = -2\kappa \frac{\partial u_{\rm ex}}{\partial \kappa},
\end{displaymath}        
\begin{displaymath}
{\mathcal C}_{\mathcal D}\Gamma\int_0^{\infty}dx x^{{\mathcal D}-2}\kappa^2 x^2\exp(-\kappa x) g(x) = 2\kappa^2 \frac{\partial^2 u_{\rm ex}}{\partial \kappa^2}.
\end{displaymath}
These expressions are exact for crystalline lattices, but remain good approximations in the strongly coupled fluid regime.
In particular, the dependence $g(x;\Gamma,\kappa)$ on $\kappa$ is known to be very weak for weakly screened ($\kappa$ is not much larger than unity) Yukawa fluids.~\cite{FaroukiJCP1994,RosenbergPRE1997,KhrapakPoP2018} The excess energy at strong coupling can be very accurately approximated as $u_{\rm ex}\simeq M_{\rm fl}\Gamma \simeq M_{\rm cr}\Gamma$, where $M_{\rm fl}$ and $M_{\rm cr}$ can be referred to as the fluid and crystalline Madelung constants ($M_{\rm fl}\sim M_{\rm cr}$).~\cite{KhrapakISM} This reflects the fact that for soft repulsive interactions the dominant contribution to the excess energy comes from static correlations.~\cite{KhrapakJCP2015} One can understand this as follows. For soft long-ranged interactions the integral in 
Eq.~(\ref{energy}) is dominated by long distances, where $g(x)$ exhibits relatively small oscillations around unity (for finite temperatures). The ratio $u_{\rm ex}/\Gamma$ is then not very sensitive to the exact shape of $g(x)$ at small $x$ (provided the correlation hole radius~\cite{KhrapakPoP2016} is properly accounted for) and, hence, to the phase state of the system.        

The consideration above implies that if $u_{\rm ex}$ (and its dependence on $\kappa$) is known, the integrals appearing in the expressions for sound velocities can be evaluated. Below we demonstrate how this works in practice in 1D, 2D, and 3D cases. 

\subsection{1D case}

The excess energy of an equidistant chain of particles is
\begin{equation}
u_{\rm ex}=\Gamma\sum_{j=1}^{\infty}\frac{e^{-\kappa j}}{j}=\Gamma\left[\kappa-\ln(e^{\kappa}-1)\right].
\end{equation}
After simple algebra we get
\begin{equation}
c_{l}^2=c_0^2\left\{\frac{\kappa e^{\kappa}[\kappa-2+2e^{\kappa}]}{(e^{\kappa}-1)^2}-2\ln(e^{\kappa}-1)\right\}.
\end{equation}
This result has been previously reported in Ref.~\onlinecite{WangPRL2001}.
It can be also obtained by direct summation
\begin{equation}
\begin{split}
c_{l}^2=c_0^2\int_0^{\infty}dxg(x)e^{-\kappa x}(2+2\kappa x+\kappa^2x^2)/x \\
=c_0^2\sum_{j=1}^{\infty}e^{-\kappa j}(2+2\kappa j+\kappa^2 j^2)/j.
\end{split}
\end{equation}
If only contribution from the two nearest neighbor particles is retained ($j=1$), the conventional dust lattice wave (DLW) sound velocity scale is obtained,~\cite{MelandsoPoP1996}
\begin{equation}\label{DLW}
c_{\rm DLW}^2=c_0^2\exp(-\kappa)(2+2\kappa+\kappa^2).
\end{equation}  

Of course, transverse mode does not exist in truly 1D case.

\begin{figure}
\includegraphics[width=7.5cm]{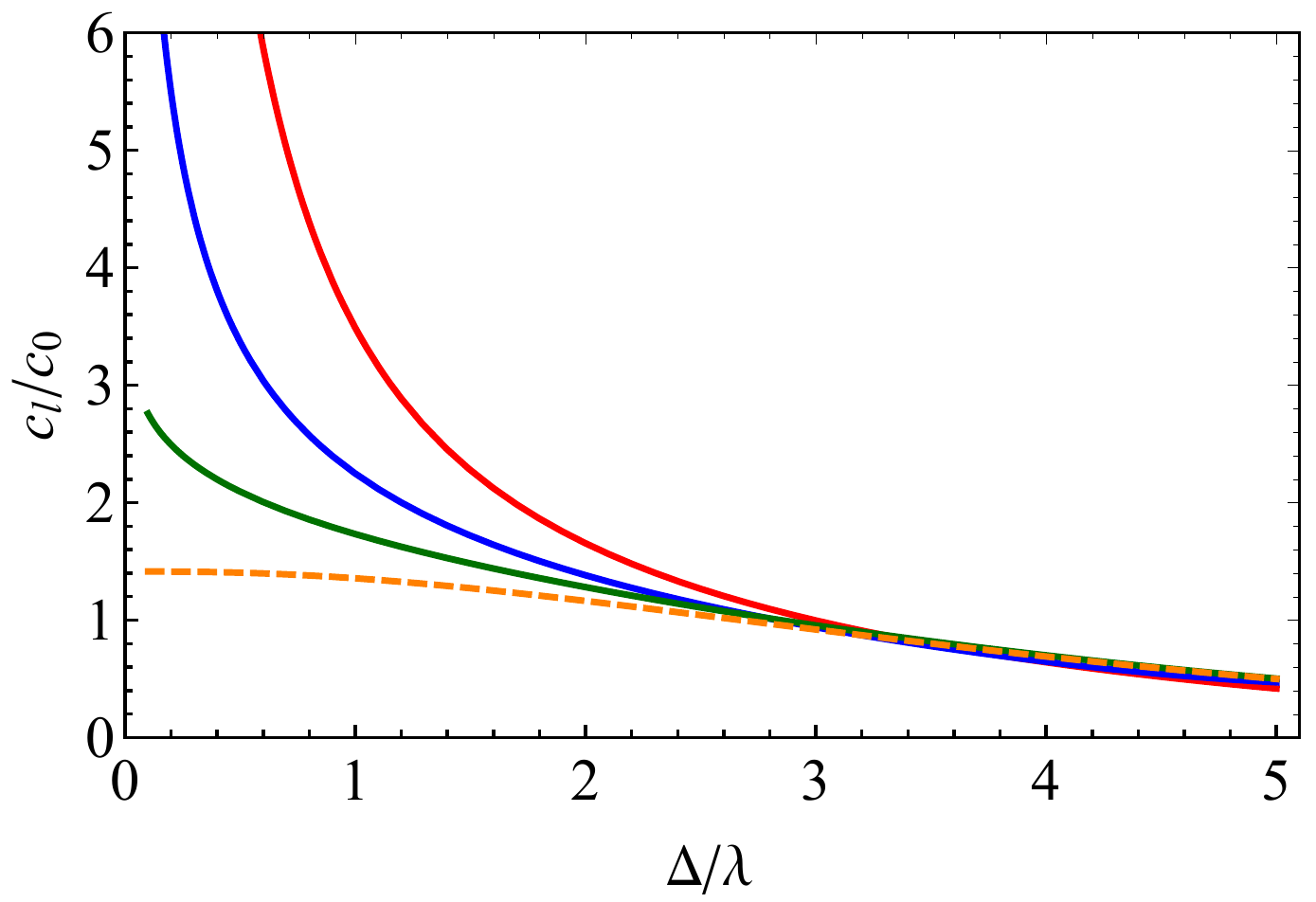}
\caption{Reduced longitudinal sound velocities versus the screening parameter $\kappa=\Delta/\lambda$ for Yukawa systems in different spatial dimensions.  The three solid curves from top to bottom correspond to 3D, 2D, and 1D cases, respectively. The dashed curve corresponds to the conventional DLW scale of Eq.~(\ref{DLW}). }
\label{Fig1}
\end{figure}

\subsection{2D case}

Combining expressions for the sound velocities and reduced excess energy and denoting $M=u_{\rm ex}/\Gamma$ we get
\begin{equation}
c_{l}^2=\frac{c_0^2}{8}\left[3\kappa^2\frac{\partial^2 M}{\partial \kappa^2}-5\kappa\frac{\partial M}{\partial \kappa}+5M\right],
\end{equation}
\begin{equation}
c_{t}^2=\frac{c_0^2}{8}\left[\kappa^2\frac{\partial^2 M}{\partial \kappa^2}+\kappa\frac{\partial M}{\partial \kappa}-M\right].
\end{equation}
The Madelung constant for the triangular lattice can be well represented by~\cite{KryuchkovJCP2017} 
\begin{equation}\label{M2D}
M=-1.9605 +0.5038\kappa -0.06236\kappa^2+0.00308\kappa^3+\frac{\pi}{\kappa}. 
\end{equation}
In Eq.~(\ref{M2D}) it is taken into account that $\kappa=\sqrt{\pi}a/\lambda$ and $\Gamma=(1/\sqrt{\pi})(Q^2/aT)$. The explicit expressions for the sound velocities are then
\begin{equation}
c_{l}^2=c_0^2\left(\frac{6.2832}{\kappa}-1.2253-0.0078\kappa^2+0.00308\kappa^3\right),
\end{equation}
\begin{equation}
c_{t}^2=c_0^2\left(0.2451-0.0234\kappa^2+0.00308\kappa^3\right).
\end{equation}
The longitudinal sound velocity diverges as $\kappa^{-1/2}$ on approaching the one-component plasma (OCP) limit, while the transverse sound velocity remains finite.

\subsection{3D case}

The relations between the longitudinal and transverse sound velocities and the Madelung constant in 3D case are
\begin{equation}
c_{l}^2=\frac{c_0^2}{15}\left[3\kappa^2\frac{\partial^2 M}{\partial \kappa^2}-4\kappa\frac{\partial M}{\partial \kappa}+4M\right],
\end{equation}
\begin{equation}
c_{t}^2=\frac{c_0^2}{15}\left[\kappa^2\frac{\partial^2 M}{\partial \kappa^2}+2\kappa\frac{\partial M}{\partial \kappa}-2M\right].
\end{equation}
The excess energy can be very well represented by the ion sphere model (ISM)~\cite{KhrapakISM,RosenfeldMolPhys1998} resulting in
\begin{equation}\label{ISM}
M=\frac{\kappa'(\kappa'+1)}{(\kappa'+1)+(\kappa'-1)e^{2\kappa'}}\left(\frac{4\pi}{3}\right)^{1/3},
\end{equation}
where $\kappa'=a/\lambda=\kappa(4\pi/3)^{-1/3}$ and the last factor in (\ref{ISM}) arises from $\Gamma=(Q^2/aT)(4\pi/3)^{-1/3}$ in the present notation.
The explicit expressions for the longitudinal and transverse sound velocities become
\begin{equation}
c_{l/t}^2=\frac{1}{15}\left(\frac{4\pi}{3}\right)^{1/3}c_0^2{\mathcal F}_{l/t}(\kappa'),
\end{equation}
where, after some algebra, we obtain 
\begin{displaymath}
{\mathcal F}_{l}(x)=\frac{x^4\left[(4+3x^2)\sinh(x)-4x\cosh(x)\right]}{\left[x\cosh(x)-\sinh(x)\right]^3},
\end{displaymath}
and
\begin{displaymath}
{\mathcal F}_{t}(x)=\frac{x^4\left[(3+x^2)\sinh(x)-3x\cosh(x)\right]}{\left[x\cosh(x)-\sinh(x)\right]^3}.
\end{displaymath}
It will be shown below that $c_l$ diverges as $\kappa^{-1}$ when the OCP limit is approached, while $c_t$ remains finite.

\begin{figure}
\includegraphics[width=7.5cm]{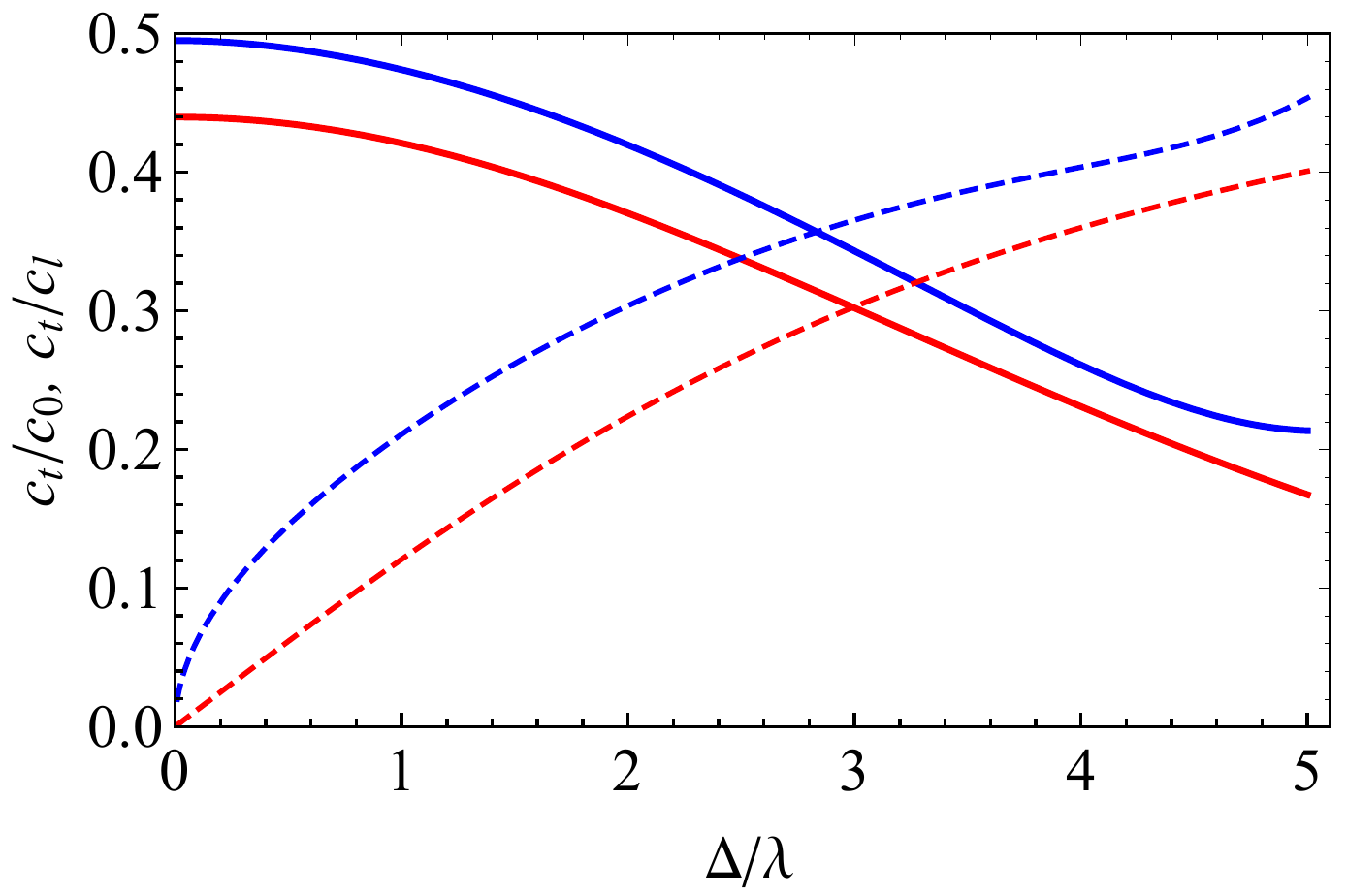}
\caption{Reduced transverse sound velocities of strongly coupled Yukawa systems versus the screenening parameter $\kappa=\Delta/\lambda$. Velocities are denoted by solid curves. Dashed curve show the ratio of longitudinal-to-transverse sound velocities. The blue (upper) curves correspond to 2D case. The red curves are for 3D case. }
\label{Fig2}
\end{figure}

\section{Main Results}\label{MainResults}

\subsection{General trends}

The calculated sound velocities are plotted in Figs.~\ref{Fig1} and \ref{Fig2}. 

Figure~\ref{Fig1} shows the longitudinal velocities for 3D, 2D, and 1D cases. In the weakly screened regime with $\kappa\lesssim 3$, the sound velocities are well separated. The highest velocity corresponds to the 3D case, the lowest one to the 1D case. Not that the sound velocities diverge as $\kappa\rightarrow 0$. This will be discussed in Sec.~\ref{OCP}. For stronger screening with $\kappa\gtrsim 3$ the longitudinal sound velocities are virtually independent of the dimensionality. They approach the common 1D DLW results with nearest neighbor interactions retained, Eq.~(\ref{DLW}). This tendency is related to the increasing steepness of the interaction potential with increasing $\kappa$. This is a general property of steep repulsive interactions, not based on the particular shape of Yukawa potential, and we will discuss this in more detail in Sec.~\ref{steep}.      

The transverse sound velocities plotted in Fig.~\ref{Fig2} are finite in the Coulomb limit and slowly decrease with increase of $\kappa$. The transverse velocity is somewhat higher in 2D than in 3D. The ratios $c_t/c_l$ start from zero at $\kappa=0$ and approach $\simeq 0.5$ as $\kappa$ increases to 5. This is yet another illustration of the strong inequality $c_l^2\gg c_t^2$ from the side of soft interactions, which has important implications in a broad physical context.~\cite{Melting2D,KhrapakMolPhys2019}        

\subsection{Weakly screened limit}\label{OCP}

In the limit of the Coulomb gas, the longitudinal dispersion relations do not exhibit acoustic asymptotes as $k\rightarrow 0$. The dispersion relation in the absence of correlations (random phase approximation) can be obtained by simply substituting $g(r)=1$ in Eq.~(\ref{omegaL}). This yields the conventional plasmon dispersion $\omega^2=\omega_p^2=4\pi Q^2n/m$ in the 3D case.  In  the 2D case the frequency grows as the square root of the wave vector, $\omega^2\propto k$. In the 1D case random phase approximation produces an integral which diverges logarithmically at small $r$. This indicates that the longitudinal sound velocities should diverge on approaching the $\kappa\rightarrow 0$ limit, as already observed. The functional for of this divergence will be established below.       

In the weakly screening limit $\kappa\ll 1$ the following series expansions of the sound velocities emerge: In 1D case we have
\begin{equation}
c_{l}=c_0\sqrt{3-2\ln \kappa};
\end{equation}
In 2D case we get
\begin{equation}
\begin{split}
c_{l}=c_0\left(\frac{2.5066}{\sqrt{\kappa}}-0.2444\sqrt{\kappa}-0.0119\kappa^{3/2}\right), \\
c_{t}=c_0\left(0.4951-0.0236\kappa^2+0.00311\kappa^3\right);
\end{split}
\end{equation}
And, finally, in 3D case the sound velocities are
\begin{equation}
\begin{split}
c_{l}=c_0\left(\frac{3.545}{\kappa}-0.0546\kappa-0.001620\kappa^{3}\right), \\
c_{t}=c_0\left(0.4398-0.0193\kappa^2+0.00055\kappa^4\right).
\end{split}
\end{equation}
Alternative fits for the sound velocities in the 3D weakly screening regime have been previously suggested in Ref.~\onlinecite{KalmanPRL2000}. 

\begin{figure}
\includegraphics[width=7.8cm]{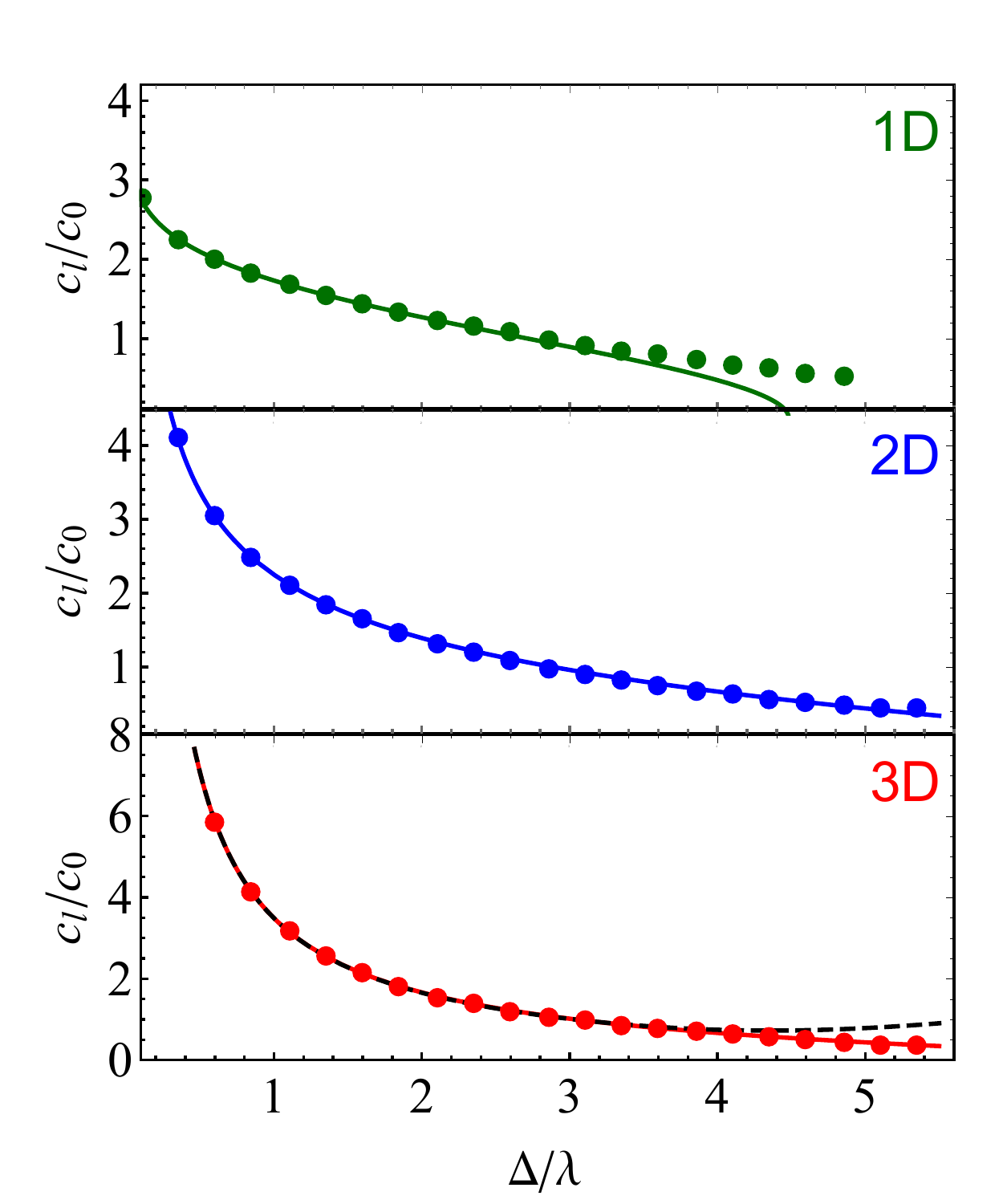}
\caption{Reduced longitudinal sound velocity versus the screening parameter $\kappa=\Delta/\lambda$. The panels from top to bottom correspond to 1D, 2D, and 3D cases, respectively. Solid curves denote the weakly screened asymptotes, symbols correspond to the full calculation. The dashed curve for the 3D case is the fit from Ref.~\onlinecite{KalmanPRL2000}. }
\label{Fig3}
\end{figure}

The weakly screened asymptotes for the longitudinal mode (solid curves) are compared with the full calculation (symbols) in Fig.~\ref{Fig3}. As the Coulomb $\kappa\rightarrow 0$ limit is approached, the longitudinal sound velocity scales as $c_{l}/c_0\sim \sqrt{-2\ln \kappa}$ (${\mathcal D}=1$), $2.5066/\sqrt{\kappa}$  (${\mathcal D}=2$), and $3.545/\kappa$  (${\mathcal D}=3$). The last two coefficients are not just fitting parameters. It is known that in the weakly screening regime (and only in this regime) the longitudinal sound velocity does not depend on the coupling strength and tends to the conventional dust acoustic wave (DAW) velocity.~\cite{RaoDAW} The details can be found in Refs.~\onlinecite{KhrapakPRE2015_Sound,KhrapakPPCF2015,SemenovPoP2015}, here we just reproduce the scalings. In the 3D case we have
\begin{equation}
c_{\rm DAW} = \omega_{p}\lambda=\sqrt{\frac{4\pi Q^2 n}{m}}\lambda=c_0\sqrt{\frac{4\pi}{\kappa^2}}\simeq c_0\frac{3.545}{\kappa}.
\end{equation}         
Similarly, in the 2D case we get~\cite{PielPoP2006}
\begin{equation}
c_{\rm DAW} = \omega_p\sqrt{\lambda}=c_0\sqrt{\frac{2\pi}{\kappa}}\simeq c_0\frac{2.5066}{\sqrt{\kappa}}.
\end{equation}

It is observed that the weakly screened asymptotes work quite well even outside the range of applicability, i.e. even at $\kappa\gtrsim 1$. The dashed curve in the bottom panel of Fig.~\ref{Fig3} corresponds to the fit proposed in Ref.~\onlinecite{KalmanPRL2000}. The agreement is excellent for $\kappa\lesssim 4$.  

\begin{figure}
\includegraphics[width=7.5cm]{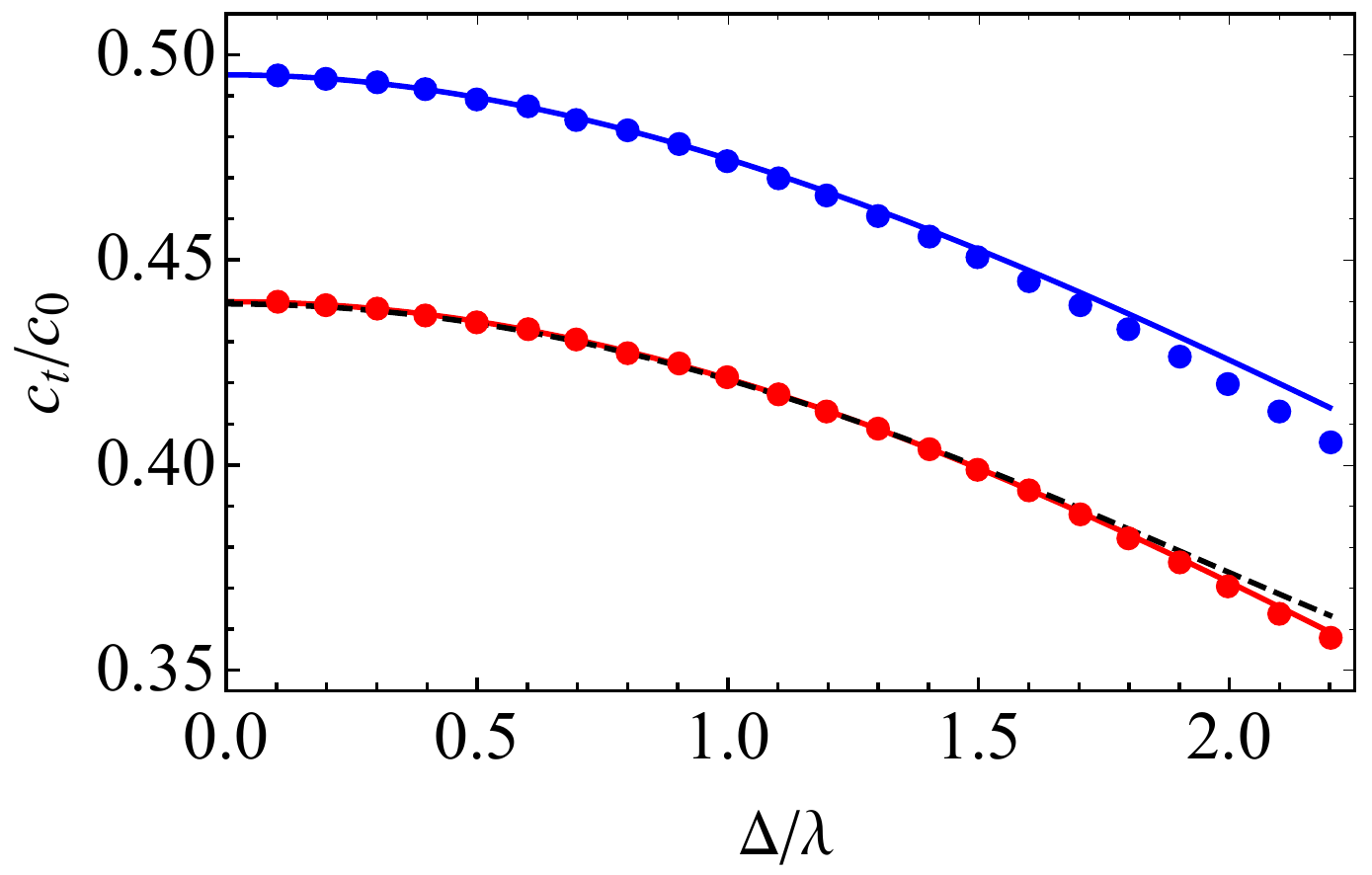}
\caption{Reduced transverse sound velocity versus the screening parameter $\kappa=\Delta/\lambda$. The top (blue) curve and symbols correspond to the 2D case, the lower (red) curves and symbols are for the 3D cases.  Solid curves denote the weakly screened asymptotes, symbols correspond to the full calculation. The dashed curve for the 3D case is the fit from Ref.~\onlinecite{KalmanPRL2000}.  }
\label{Fig4}
\end{figure}

The results for the transverse sound velocity in 2D and 3D are plotted in Figure~\ref{Fig4}. The solid curves denote the weakly screened asymptotes, symbols correspond to the full calculation, the dashed curve is the 3D fit from Ref.~\onlinecite{KalmanPRL2000}. We observe that the weakly screened asymptotes are appropriate only for $\kappa\lesssim 2$ in this case. The transverse velocities do not vary much in the considered range of $\kappa$ and remain finite in the limit $\kappa\rightarrow 0$. We have $c_{t}/c_0\simeq 0.495$ (${\mathcal D}=2$) and $0.440$ (${\mathcal D}=3$). How this compares with the known results for the one-component plasma (OCP) systems with Coulomb interactions in 2D and 3D?  For the OCP systems the transverse sound velocities are directly related to the thermal velocity and the reduced excess energy.~\cite{GoldenPoP2000} In the 2D case we have
\begin{displaymath}
c_{t}^2=-\frac{1}{8}v_{\rm T}^2 u_{\rm ex}.
\end{displaymath}
Combining this with the strong coupling asymptote,~\cite{KhrapakCPP2016} $u_{\rm ex}\simeq -1.106103(Q^2/aT)$, we get $c_{t}/c_0\simeq 0.495$, in excellent agreement with the result above. Similarly, in the 3D case we have  
\begin{displaymath}
c_{t}^2=-\frac{2}{15}v_{\rm T}^2 u_{\rm ex}.
\end{displaymath}  
Using the ISM estimation of the OCP excess energy,~\cite{KhrapakPoP2014,DubinRMP1999} $u_{\rm ex}\simeq -\tfrac{1}{9} (Q^2/aT)$ we get $c_{t}/c_0\simeq 0.440$, again in excellent agreement with the result above. The dashed curve in the 3D case corresponds to the fit from Ref.~\onlinecite{KalmanPRL2000}. For $\kappa\lesssim 2$ all the data shown are almost coinciding. 

\subsection{Sound velocities for steep repulsive potentials}\label{steep}

For steep repulsive potentials we should have $|f'(x)/x|\ll |f''(x)|$. Then the main contribution to the sound velocities comes from the second derivative of the potential. This main contribution to the longitudinal sound velocity can be evaluated from
\begin{equation}
c_{l}^2=c_0{\mathcal B}_l{\mathcal C}_{\mathcal D}\int_0^{\infty}dx x^{{\mathcal D}+1} g(x) f''(x), 
\end{equation}  
where as usually in this paper $x=r/\Delta$. Further, for steep interactions the main contribution to the integral above comes from the first shell of neighbors at $x\simeq 1$. We can therefore approximate $x^2f''(x)$ by $f''(1)$ under the integral. Such substitution is exact only for a long-range logarithmic potential, but should provide a good estimate for quickly decaying potentials and an RDF $g(x)$ that has a strong peak near $x\simeq 1$. The remaining of the integral can be related to the number of nearest neighbors $N_{\rm nn}$ using
\begin{equation}
\begin{split}
{\mathcal C}_{\mathcal D}\int_0^{\infty}x^{{\mathcal D}+1}g(x)f''(x)dx\simeq \\ {\mathcal C}_{\mathcal D}\int_0^{x_{\min}}x^{{\mathcal D}-1}g(x)f''(1)dx\simeq f''(1)N_{\rm nn},
\end{split}
\end{equation} 
where $x_{\min}>1$ is roughly the position of the first non-zero minimum of $g(x)$ (in the considered situation the value of the integral is not sensitive to $x_{\min}$, because the main contribution comes from the immediate vicinity of $x=1$). Taking into account that at strong coupling $N_{\rm nn}\simeq 12$ (${\mathcal D}=3$),  6 (${\mathcal D}=2$), and 2 (${\mathcal D}=1$), we get 
\begin{equation}
\begin{split}
c_{l}^2=\frac{\epsilon}{m}f''(1), \quad\quad ({\rm 1D}) \\
c_{l}^2=\frac{18}{16}\frac{\epsilon}{m}f''(1), \quad\quad ({\rm 2D}) \\
c_{l}^2=\frac{12}{10}\frac{\epsilon}{m}f''(1). \quad\quad ({\rm 3D}) 
\end{split}
\end{equation}  
Thus the, longitudinal sound velocities are all proportional to $\sqrt{(\epsilon/m)f''(1)}$, multiplied by a coefficient of order unity. This coefficient has the following scaling with the dimensionality: 3D:2D:1D$\simeq \sqrt{1.2}:\sqrt{1.13}:1$. The difference in the coefficients is insignificant taking into account simplifications involved. This explains, why all the curves approach the common asymptote as $\kappa$ increases in Fig.~\ref{Fig1}. This common asymptote is just the DLW nearest neighbor result of Eq.~(\ref{DLW}).

Note that within this approximation the ratio of the longitudinal to transverse sound velocities is $c_{t}/c_{l}=1/\sqrt{3}\simeq 0.58$, independently of dimensionality. The dashed curves in Fig.~\ref{Fig2} should approach this asymptote as $\kappa$ increases further. Note, however, that the QCA approach itself cannot be applied for arbitrary large $\kappa$. It loses its applicability when approaching the hard sphere interaction limit.~\cite{KhrapakJCP2016,KhrapakSciRep2017}    

In the Appendix~\ref{freezing} we discuss how the consideration in this Section can lead to a simple freezing indicator, which was previously applied to various classical 3D fluids and, particularly successfully, to the 3D Yukawa fluid.

\section{Conclusion}\label{Concl}

The effect of spatial dimensions on the amplitude of sound velocities in strongly coupled Yukawa systems has been investigated. A unified approach, based on infinite frequency (instantaneous) elastic moduli of fluids and isotropic solids has been formulated. In this approach, the sound velocities are expressed in terms of the excess internal energy, which is very well known quantity for Yukawa systems. Physically motivated expressions, convenient for practical application have been derived and analyzed. Relations to dust-acoustic wave (DAW) and dust-lattice wave (DLW) velocities have been explored. The regimes of weak and strong screening have been analyzed separately. It has been demonstrated that at weak screening ($\kappa\lesssim 3$) the longitudinal sound velocities in different spatial dimensions are well separated and their amplitude increases with dimensionality. For stronger screening ($\kappa\gtrsim 3$), the longitudinal sound velocities in different dimensions all approach the same DLW asymptote, and this can be a very useful observation for practical applications. The explanation of this tendency has been provided.

\begin{acknowledgments} 
I would like to thank Viktoria Yaroshenko for reading the manuscript.
\end{acknowledgments}

\appendix*

\section{Related freezing indicator}\label{freezing}

To the same level of accuracy as in Sec.~\ref{steep} we can estimate the Einstein frequency in 3D systems with steep interparticle interactions as
\begin{equation}\label{efreq}
\Omega_{\rm E}^2 = \frac{n}{3m}\int d{\bf r}\Delta\phi(r) g(r) \simeq \frac{\epsilon N_{\rm nn}}{3m\Delta^2}f''(1)\propto \frac{\phi''(\Delta)}{m}. 
\end{equation} 
The celebrated Lindemann melting criterion~\cite{Lindemann} states that melting occurs when the particle root-mean-square vibrational amplitude around the equilibrium position reaches a threshold value of about $0.1$ of the interparticle distance. Its simplest version (assuming the Einstein approximation for particle vibrations in the solid state) may be cast in the form
\begin{equation}\label{Lindemann}
\langle\xi^2\rangle \simeq \frac{3T}{m\Omega_{\rm E}^2} \simeq L^2\Delta^2,
\end{equation}   
where $L$ is the Lindemann parameter. Combining Eqs.~(\ref{efreq}) and (\ref{Lindemann}) we immediately see that at the fluid-solid phase transition one may expect
\begin{equation}\label{indicator}
\frac{\phi''(\Delta)\Delta^2}{T}\simeq {\rm const}.
\end{equation}
This kind of criterion was first applied to Yukawa systems,~\cite{VaulinaJETP2000,VaulinaPRE2002,FortovPRL2003} in which case it works very well for $\kappa\lesssim 5$. It was also applied with some success to Lennard-Jones (LJ) systems~\cite{KhrapakPRB2010,KhrapakMorfJCP2011} and LJ-type systems,~\cite{KhrapakJCP2011} where it is able to approximately predict the liquid boundary of the liquid-solid coexistence region (freezing transition). For potentials, exhibiting anomalous re-entrant melting behavior, such as the exp-6 and Gaussian Core Model, the agreement with numerical data is merely qualitative and its application is limited to the low-density region.~\cite{KhrapakMolPhys2011} From the derivation, it is expected that the freezing indicator (\ref{indicator}) is more appropriate for steep interactions. Why it works so well for soft weakly screened Yukawa systems (including OCP), remains to some extent mysterious. Note, however, that an alternative derivation of the freezing indicator (\ref{indicator}) for Yukawa systems, based on the isomorph theory approach, has been recently discussed.~\cite{VeldhorstPoP2015}     

Application of this freezing indicator to 2D and 1D systems is not possible in view of the predicted divergence of $\langle \xi^2 \rangle$ in these spatial dimensions due to long-wavelengths density fluctuations.~\cite{Landau1937,JancoviciPRL1967}  

\bibliographystyle{aipnum4-1}
\bibliography{SoundDim_References}

\end{document}